# Anomalous decrease of relatively large shocks and increase of the *p* and *b* values preceding the April 16, 2016 *M*7.3 earthquake in Kumamoto, Japan

Author #1: K. Z. Nanjo, Global Center for Asian and Regional Research, University of Shizuoka, 3-6-1, Takajo, Aoi, Shizuoka 420-0839, Japan, Tel: +81-54-245-5600, Fax: +81-54-245-5603, E-mail: nanjo@u-shizuoka-ken.ac.jp

Author #2: A. Yoshida, Center for Integrated Research and Education of Natural Hazards, Shizuoka University, 836, Oya, Suruga, Shizuoka 422-8529, Japan, Tel: +81-54-238-4502, Fax: +81-54-238-4911, E-mail: akio.yoshi@nifty.com

The corresponding author: K. Z. Nanjo

**Abstract**

The 2016 Kumamoto earthquakes in Kyushu, Japan, started with a magnitude (*M*) 6.5 quake on April 14 on the Hinagu fault zone (FZ), followed by active seismicity including an *M*6.4 quake. Eventually, an *M*7.3 quake occurred on April 16 on the Futagawa FZ. We investigated if any sign indicative of the *M*7.3 quake could be found in the space-time changes of seismicity after the *M*6.5 quake. As a quality control, we determined in advance the threshold magnitude, above which all earthquakes are completely recorded. We then showed that the occurrence rate of relatively large ($M \geq 3$) earthquakes significantly decreased 1 day before the $M$7.3 quake. Significance of this decrease was evaluated by one standard deviation of sampled changes in the rate of occurrence. We next confirmed that seismicity with $M \geq 3$ was well modeled by the Omori-Utsu (OU) law with $p \sim 1.5 \pm 0.3$, which indicates that the temporal decay of seismicity was significantly faster than a typical decay with $p = 1$. The larger $p$ value was obtained when we used data of the longer time period in the analysis. This significance was confirmed by a bootstrapping approach. Our detailed analysis shows that the large $p$ value was caused by the rapid decay of the seismicity in the northern area around the Futagawa FZ. Application of the slope (the $b$ value) of the Gutenberg–Richter (GR) frequency-magnitude distribution to the spatiotemporal change in the seismicity revealed that the $b$ value in the northern area increased significantly, the increase being $\Delta b = 0.3$-$0.5$. Significance was verified by a statistical test of $\Delta b$ and a test using bootstrapping errors. Based on our findings, combined with the results obtained by a stress inversion analysis performed by the National Research Institute for Earth Science and Disaster Resilience, we suggested that stress near the Futagawa FZ had reduced just prior to the occurrence of the $M$7.3 quake. We proposed, with some other observations, that a reduction in stress might have been induced by growth of the slow slips on the Futagawa FZ.

**Introduction**

Seismicity has become high in almost all parts of Japan since the 2011 magnitude ($M$)

9.0 Tohoku-Oki earthquake (e.g., Ishibe et al., 2011; Toda et al., 2011). The 2016 Kumamoto earthquakes occurred under these circumstances, leading to a memorable event that once again caused devastating damage in Japan. The Kumamoto earthquakes in the Kyusyu district began with an $M$6.5 quake on April 14 2016, at 21:26, on the Hinagu fault zone (FZ), which was followed by numerous shocks including an $M$6.4 quake on April 15, at 00:03. Eventually, on April 16, at 01:25, an $M$7.3 quake occurred on the Futagawa FZ. The probability of the occurrence of an $M$7.0 or so earthquake on the Futagawa FZ had been estimated by the ERC to be 0~0.9% within 30 years (Earthquake Research Committee, 2016a), which was rather high among the probability at active faults in Japan. In a press conference that was held just after the $M$6.5 quake, the Japan Meteorological Agency (JMA) warned of the possibility of large aftershocks that should bring further damages. However, no information on an increased probability of $M$7 or larger quakes was announced by the JMA during the period between the $M$6.5 quake and the $M$7.3 quake. This is because, according to the ERC (1998) protocol (see also ERC, 2016b), the JMA had not considered the occurrence of larger earthquakes after an $M$6.5 quake. Moreover, it usually takes more than one day after a mainshock occurrence for the JMA to issue an announcement of the probability of large aftershocks.

The objective of this study was to investigate if any indication of the occurrence of the $M$7.3 quake had not appeared in the space-time changes of the rate of occurrence and magnitude-number distribution of shocks after the $M$6.5 quake on April 14. We found that $M$3 or larger shocks decreased during the period of one day preceding the $M$7.3 quake and that the $b$ value (the ratio of small to large events) became pronouncedly high, corresponding to a decrease of comparatively large shocks, especially in the area near the hypocenter of the $M$7.3 quake.

Tremendous efforts have been made to find effective signatures that indicate a large earthquake occurrence in the near future. A temporal decrease in the *b* value may be a candidate (e.g. Suyehiro et al., 1964; Nanjo et al., 2012) while another may be seismic quiescence (e.g., Wiemer and Wyss, 1994; Sobolev and Tyupkin, 1997). It is known that aftershock activity becomes significantly quiescent preceding large aftershocks (Mtasu'ura, 1986). However, any signatures ever proposed do not seem to be universal, since accelerating seismicity (e.g., Bowman et al., 1998) and an increase in the *b* value (Smith, 1981) were occasionally reported.

We examined the spatial and temporal trends in the rate of occurrence of comparatively large ($M \geq 3$) shocks and the *b* value in the seismic activity following the *M*6.5 quake on April 14. We found that the *b* value in the activity in the northern area became significantly high before the *M*7.3 quake on April 16, and that this feature corresponded to a decrease of large ($M \geq 3$) shocks. We consider that the high *b* value and the decrease of comparatively large shocks were precursors to the *M*7.3 quake and propose what caused the observed changes in the space-time seismicity.

**Methods and data**

Activity following the *M*6.5 quake is well modeled by two statistical laws: the Gutenberg–Richter (GR) frequency–magnitude law (Gutenberg and Richter, 1944) and the Omori–Utsu (OU) aftershock decay law (Utsu, 1961).

The GR law is given as $\log_{10}N = a - bM$, where *N* is the number of earthquakes per unit time with magnitudes greater than or equal to *M*, *a* describes the productivity of the regional seismicity, and the *b* value is the ratio of small to large events. A high *b* value indicates a larger proportion of small earthquakes, and vice versa. In the laboratory and in the Earth's crust, the *b* value is known to be inversely dependent on differential stress (Scholz, 1968, 2015). In this context, measurements of spatial temporal changes in the *b* value could be used to discover highly stressed areas where future ruptures are likely to occur (Schorlemmer and Wiemer, 2005, Nanjo et al., 2012). To consistently estimate *b* values over space and time, we employed the EMR (Entire-Magnitude-Range) technique (Woessner and Wiemer, 2005), which also simultaneously calculates the completeness magnitude $M_c$, above which all events are considered to be detected by a seismic network. EMR applies the maximum-likelihood method (Aki, 1965) in computing the *b* value to events with magnitudes above $M_c$. Uncertainties in *b* and $M_c$ were computed by bootstrapping (Schorlemmer et al., 2003, 2004).

Any completeness estimation method is subject to errors, particularly during productive seismicity times such as investigated in this study. To confirm implications of this study, we compared $M_c$ by the EMR method with that by other three methods. Results of the comparison are shown in Appendix 1 and Fig. A1 in Additional file 1.

The OU law is given as $\lambda = k/(c + t)^{-p}$, where *t* is the time since a mainshock occurrence, $\lambda$ is the number of aftershocks per unit time at *t* with magnitudes greater than or equal to a cut-off *M*, and *c*, *k*, and *p* are constants. $p = 1$ is generally a good approximation (Omori, 1894). From the view point of the rate- and state-dependent friction law (Dieterich 1994), variability in *p* can be observed: $p > 1$ and $p < 1$ for rapidly and slowly decreasing stress, respectively. When stress increases with time, $p = 1$ at $t \gg 0$. If this friction law is valid in the Earth's crust, then the *p* values could be used to infer stress history. Similar to the GR case, we used a maximum-likelihood fit to determine the parameters for this law. Uncertainties in the *p* value were computed by bootstrapping.

Our dataset is the earthquake catalog maintained by JMA. From this catalog, we separated events associated with the 2016 Kumamoto earthquakes after the occurrence of the $M$6.5 quake on April 14. Owing to a large increase in the number of borehole seismic stations (Obara et al., 2005), event delectability has been greatly improved, which lowers the minimum magnitude for catalog homogeneity (Nanjo et al., 2010). Since 2000, the minimum magnitude is about $M = 0 \sim 1$ in the Kyushu district.

Small events in clusters such as swarms and aftershocks are often missed in the earthquake catalog, as they are ''masked'' by the coda of large events and overlap with each other on seismograms. According to previous cases (e.g., Helmstetter et al., 2006; Nanjo et al., 2007), $M_c$ depends on time $t$. In creating Fig. 1, we used a moving window approach, whereby the window covered 200 events (red). $M_c$ decreased with $t$ from 2.9 and reached a constant value at around 2.2. Relatively large events occurred early in the sequence, and the mean magnitude of these events evolved to small values over time (grey). The time-dependent decrease of $M_c$ is consistent with the data. To examine if the occurrence of the $M$6.4 quake affected completeness, events before the $M$6.4 quake were excluded from the dataset and $M_c$ was assigned as a function of $t$ since the $M$6.4 quake (blue). Immediately after the $M$6.4 quake, the estimate of $M_c$ was as high as 2.4. $M_c$ decreased with $t$ and reached a constant value at around 2.2. We were unable to see any significant deviation of the $M_c$-$t$ pattern since the $M$6.4 quake (blue) from since the $M$6.5 quake (red). This is interpreted as an indication that the effect of the $M$6.4 quake on $M_c$ was weak.

**Results**

As a preliminary analysis to modeling the OU law, the cumulative number of $M \geq 3$ earthquakes was calculated as a function of relative time to the $M$7.3 quake over the entire period of activity (Fig. 2a). Taking $M_c$ in Fig. 1 into consideration, we truncated the catalog at $M = 3$ in advance to discard all data that may have been inhomogeneous. Visual inspection of the cumulative curve indicates that there are two kinks at the times 0.98 and 0.67 days before the $M$7.3 quake, suggesting that transitions to lower occurrence rates of $M \geq 3$ shocks happened twice. We divided all shocks into two groups: shocks in the northern area (higher than 32.72° in latitude) and those in the southern area (lower than 32.72° in latitude). We employed the same plotting procedure as for Fig. 2a to show that there were again two kinks at 0.98 and 0.67 days before the $M$7.3 quake in both groups (Figs. 2d and 2g). Figs. 2d and 2g show that the timing of the kinks did not correspond to the occurrence of the $M$6.4 quake (star at 1.05 days preceding the $M$7.3 quake), indicating that this quake did not affect the seismic activity as a whole.

We also checked the change in the number (i.e., the change in the rate of occurrence) of $M \geq 3$ shocks. To do this, we first plotted the number (the rate of occurrence) of $M \geq 3$ shocks in the entire area as a function of relative time to the $M$7.3 quake in Fig. 2b, where we counted $M \geq 3$ shocks in each time window of 0.08 days. We then plotted the change in the number (the change in the rate of occurrence) of those shocks in the same area as a function of relative time to the $M$7.3 quake in Fig. 2c. Also included in this figure are the mean and standard deviation of the change in the number of $M \geq 3$ shocks. At 0.98 days relative to the $M$7.3 quake, the change in the number was negative and smaller than the standard deviation, while the change at 0.67 days before the $M$7.3 quake was also at negative but within the margins of the standard deviation. This shows that the transition to a lower occurrence rate at 0.98 days relative to the $M$7.3 quake was significant, but the transition at 0.67 days might not have been. The analysis was performed for the northern area (Figs. 2e-f) and southern areas (Figs. 2h-i) separately, obtaining the same feature. We confirmed that the above-described result is robust, by sampling different time windows (0.10 and 0.05 days) and by noting that the general features are the same for all the cases (Fig. A2 in Additional file 1). A preliminary analysis of the change in the number of $M \geq 3$ shocks suggests that the decay of seismic activity over time was faster than that predicted by the OU law, typically with $p = 1$.

Fig. 3 shows a good fit of the OU law with $p = 1.49 \pm 0.29$ to the activity with $M \geq 3$ for the entire area during the period between the $M6.5$ and $M7.3$ quakes ($t = 0$-1.16 days). Note that this value is significantly larger than a typical value ($p = 1$). For a short time period ($t = 0$-0.18 days, equivalent to 1.16-0.98 days preceding the $M7.3$ quake), $p = 0.46 \pm 0.14$ is significantly below 1. For an intermediate period ($t = 0$-0.49 days, corresponding to 1.16-0.67 days preceding the $M7.3$ quake), $p = 0.99 \pm 0.26$ lies between these two $p$ values. The difference in the $p$ value between the short and long periods is statistically significant, showing that decay of seismic activity became faster as period increased, as expected from a preliminary analysis. The influence of period on $p$ is shown in the inset of Fig. 3 (data in black). We conducted the same analysis for shocks that occurred in the northern and southern areas. The results are plotted in the inset of Fig. 3 (data in cyan and pink, respectively). The difference in the $p$ value between these areas is significant for periods longer than 0.5 days, where the $p$ value was not plotted for the southern area for periods longer than one day because the model-fitting analysis did not converge. It can be said that the rapid decrease in the number of $M \geq 3$ shocks in the northern area was responsible to the rapid decay in the activity observed over the entire area.

We next examined the spatial distributions in the $b$ value for two periods, 1.16-0.98 and 0.98-0 days prior to the $M7.3$ quake. The results are shown in Figs. 4a and 4d, respectively. To create these figures, we calculated $b$ values for events falling within a cylindrical volume with a 5 km radius, centered at each node on a 0.005º × 0.005º grid and plotted a $b$ value at the corresponding node if at least 20 events in the cylinder yielded a good fit to the GR law. We assumed that the characteristic dimension of the node spacing (~500 m) was a larger value than the value of location uncertainty (typically, 275 m) of the JMA catalog (Richards-Dinger et al., 2010). Distributions in the $M_c$ value that were simultaneously calculated with a $b$ value at each node are shown in Figs. 4b and 4e. Overall, the $b$ values increased with time. The difference $\Delta b = b_{1.16-0.98} - b_{0.98-0}$ is mostly above 0 (cyan to blue colors in Fig. 4g), and ranges from $\Delta b = -0.05$ to $\Delta b = 0.53$. A map view shows a zone of large increase in the $b$ value ($\Delta b \geq 0.3$, blue colors), but only in the northern area. Observed changes in the $b$ value are not considered significant if the test proposed by Utsu (1992, 1999) is not passed, as is shown in Fig. 4h. If $\log P_b$, the logarithm of the probability that the $b$ values are not different, is equal to or smaller than -1.3 ($\log P_b \leq -1.3$), then the change in $b$ is significant (Schorlemmer et al., 2004). A map of $\log P_b$ (Fig. 4h) revealed a zone of the low values ($\log P_b \leq -1.3$, green colors) in the northern area. This observation is further supported by applying a bootstrapping approach (Schorlemmer et al., 2003, 2004) to the standard deviation of the $b$ values, σ. The average standard deviation $\bar{\sigma}$ of the early period (1.16-0.98 days prior to the $M$7.3 quake) is 0.08 and 90% of the σ values are smaller than 0.10 (Fig. 4c). For the later period (0.98-0 days prior to the $M$7.3 quake), $\bar{\sigma}$ is 0.12 and 90% of the values are smaller than 0.18. Higher errors of $\sigma \geq 0.18$ (yellow) were obtained only in the northern area. The sums between σ in the early period and σ in the late period are at most 0.3. Comparing $\Delta b$ with σ, we can say that the increase in the $b$ value in part of the northern area is significant. We conducted the same statistical test as shown in for Fig. 4, but calculated $b$, $M_c$, and σ values, by sampling the nearest 100 earthquakes at each node (Fig. A3 in Additional file 1). The general feature remained similar to Fig. 4. Our statistical test demonstrates that the observed changes in the *b* value were physically meaningful, and not caused by artifacts.

If the *b* value offers an indicator of stress in the Earth's crust, then it is expected to have increased by the time the *M*7.3 quake occurred. To examine this possibility, we compared the *b* values before the *M*6.5 quake with the *b* values after the *M*7.3 quake in a polygon that includes epicenters of the *M*6.5, *M*6.4, and *M*7.3 quakes (Fig. A4 in Additional file 1). The *b* value was almost stable when *b* was 0.6~0.8 before the *M*6.5 quake, consistent with the results obtained by Nanjo et al. (2016). We found that though the *b* value immediately after the *M*7.3 quake fluctuated due to many aftershocks following the quake, that the *b* values after this period are significantly larger than those before the *M*6.5 quake. This result indicates that measurement of spatial and temporal changes in the *b* value is useful to infer stress in the Earth's crust, supporting our interpretation that the significant increase in the *b* value in Fig 4 shows a decrease in stress before the *M*7.3 quake.

**Discussion**

We found that the rate of occurrence of comparatively large shocks with *M*3 or larger in the seismic activity after the *M*6.5 quake on April 14, 2016 decreased significantly preceding the *M*7.3 quake on April 16. We further demonstrated that the *p* and *b* values increased significantly before the *M*7.3 quake, a feature that is consistent with the decrease of comparatively large shocks. Our analysis of the spatial and temporal changes in the *p* and *b* values indicates that the increase of the *p* and *b* values was most conspicuous in the northern area of the seismic activity after the *M*6.5 quake. This finding suggests that stress around the Futagawa FZ had begun to decrease preceding the *M*7.3 quake, if we considered that a high *p* value declines an increase in stress (Dieterich, 1994) and that a high *b* value indicates low stress (Scholz, 1968, 2015).

The reduction in the normal stress at the Futagawa FZ, i.e., a decrease of compressive stress on the fault, is suggested by an independent observation from the NIED, the National Research Institute for Earth Science and Disaster Resilience (2016). NIED examined the temporal change in the stress field by performing a stress inversion analysis using focal mechanisms of earthquakes in and around the Kumamoto earthquakes. The minimum compressive stress ($\sigma_3$) axis lay roughly in a north-south axis before the *M*6.5 quake. A comparison between the $\sigma_3$ axis after the *M*6.5 quake and the $\sigma_3$ axis before this quake shows that the axis rotated counterclockwise and became nearly perpendicular to the strike of the Futagawa FZ, which indicates the normal force operating to the fault plane reduced.

In the seismic activity after the *M*6.5 quake, we found that earthquakes had occurred near the hypocenter of the *M*7.3 quake in addition to the majority of the seismicity around the Hinagu FZ (Fig. 5a). Fig. 5b shows a 5 km wide cross-section nearly perpendicular to the strike of the Futagawa FZ, on which the rupture of the *M*7.3 quake occurred (ERC, 2016a). Note that several events are recognized close to the hypocenter of the *M*7.3 quake. In the magnitude-time diagram (Fig. 5c) where these events are designated by bold squares, they occurred 0.98 days before the *M*7.3 event. One possible explanation may be that they were produced by a nucleation process, as suggested by Ohnaka (1993), of the *M*7.3 quake. Our finding of the occurrence of earthquakes close to the hypocenter of the *M*7.3 quake (Fig. 5), combined with the reduction of normal pressure on the Futagawa FZ (NIED, 2016), indicates growth of a quasi-static pre-slips before the *M*7.3 quake. The pre-slips might have relaxed stress around the hypocenter of the *M*7.3 quake. This stress relaxation provided feedback to reduce the occurrence rate of larger ($M \geq 3$) shocks (Fig. 2), causing a significantly larger $p$ value than $p = 1$ (Fig. 3), and a significant increase in the $b$ value (Fig. 4).

**Additional file**

Additional file 1. This file includes Appendix 1, Figures A1-A4, and Additional references.

**List of abbreviations**

EMR method: Entire-Magnitude-Range method, ERC: Earthquake Research Committee, FZ: fault zone, GR law: Gutenberg–Richter law, JMA: Japan Meteorological Agency, NIED: National Research Institute for Earth Science and Disaster Resilience, OU law: Omori-Utsu law.

**Competing interests**

The authors declare no competing financial interests.

**Authors' contributions**

KZN performed numerical simulations, analyzed data and prepare the figures. KZN and AY helped to draft the manuscript and participated in interpretation. KZN wrote the final manuscript. Both authors read and approved the final manuscript.

**Acknowledgements**

We thank the Editor (Matha Savage) and two anonymous reviewers for their constructive review.

**Figure legends**

**Fig. 1.** Plot of $M_c$ as a function of relative time to the $M$7.3 quake (vertical solid line) since the $M$6.5 quake (red) and since the $M$6.4 quake (blue) in which a 200-event window was used. Uncertainty was according to the bootstrapping (Schorlemmer et al., 2003, 2004). Also included in this figure is a $M$-time diagram. Vertical dashed lines indicate the relative times, 0.98 and 0.67 days, to the $M$7.3 quake. Red stars: $M$6-class quakes ($M$6.5, $M$6.4), yellow star: $M$7.3 quake.

**Fig. 2.** Plots of the cumulative number (**a**, **d**, **g**), the number (i.e., the rate of occurrence) (**b**, **e**, **h**), and the change in the number (i.e., the change in the rate of occurrence) (**c**, **f**, **i**) of $M \geq 3$ earthquakes as a function of relative time (days) to the $M$7.3 quake (vertical solid line). **a-c** seismicity in the entire area, **d-f** seismicity in the northern area (higher than 32.72° in latitude), and **g-i** seismicity in the southern area (lower than 32.72° in latitude). We used a time window of 0.08 days. See also Fig. A2 in Additional file 1 for other time windows of 0.05 and 0.10 days. Vertical dashed lines indicate relative times 0.98 and 0.67 days to the $M$7.3 quake.

**Fig. 3.** Number $\lambda$ (day$^{-1}$) of $M \geq 3$ earthquakes in the entire area as a function of $t$ (days) from the $M$6.5 quake. Fitting of the OU law to seismicity in the periods $t$ = 0-0.18 (green), 0-0.49 (blue), and 0-1.16 (red) results in $(p, c)$ = (0.46 ± 0.14, 0.00 ± 0.00), (0.99 ± 0.26, 0.02 ± 0.02), and (1.49 ± 0.29, 0.06 ± 0.04), respectively, where $c$ (days) is a constant of the OU law. Inset: $p$ value as a function of the length of the analyzed period (days) since the $M$6.5 quake for seismicity in the entire area (black), in the northern area (cyan), and in the southern area (pink).

**Fig. 4**. Results from a test of significance of the difference between the periods 1.16-0.98 days and 0.98-0 days prior to the $M$7.3 quake. **a** $b$ values, **b** $M_c$ values, **c** σ values from 1.16-0.98 days. **d** $b$ values, **e** $M_c$ values, **f** σ values from 0.98-0 days. **G** Δ$b$, the difference in $b$ values between the periods 1.16-0.98 and 0.98-0 days. **h** log $P_b$, the logarithm of the probability that the $b$ value for 1.16-0.98 days is different from the $b$ value for 0.98-0 days. Red stars: $M$6-class quakes ($M$6.5, $M$6.4), yellow star: $M$7.3 quake, red lines: surface traces of active faults, where Hinagu FZ and Futagawa FZ are designated.

**Fig. 5. a** The epicentral map of earthquakes that occurred during the period from the $M$6.5 quake to the $M$7.3 quake. Events marked by dark grey squares were used to create the cross-sectional view in **b** and the magnitude-time diagram in **c**. Red stars: $M$6-class quakes ($M$6.5, $M$6.4), yellow star: $M$7.3

quake, red lines: surface traces of active faults, where Hinagu FZ and Futagawa FZ are designated. **b** Cross-sectional view of the seismicity. Apart from the majority of the events, outliers (bold squares) can also be observed around the hypocenter of the $M$7.3 quake. **c** $M$ as a function of the time (days) until the $M$7.3 quake (vertical solid line). Outliers seen in **b** are designated by bold squares. Vertical dashed lines: 0.98 days and 0.67 days before the $M$7.3 quake.

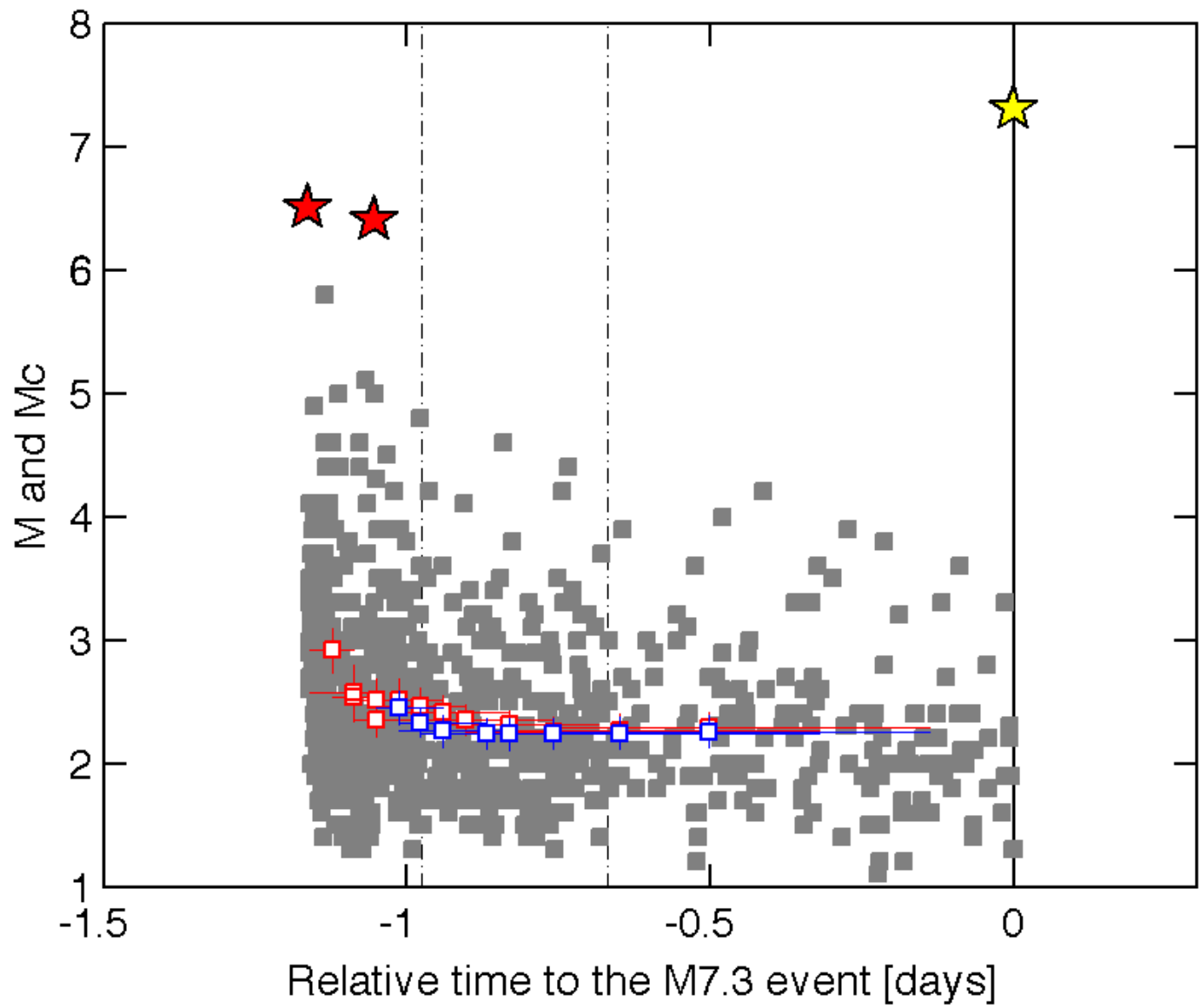

Fig. 1

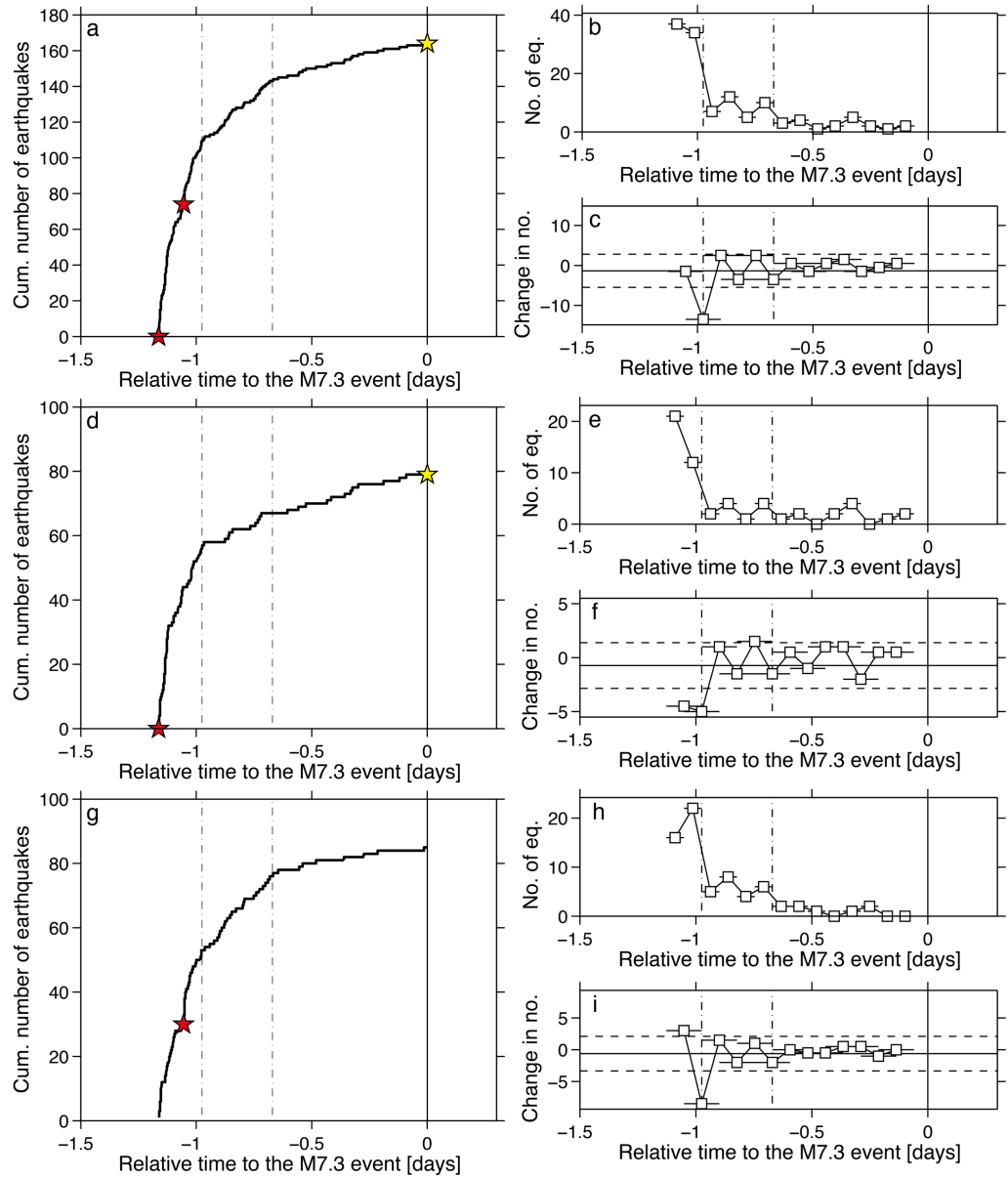

Fig. 2

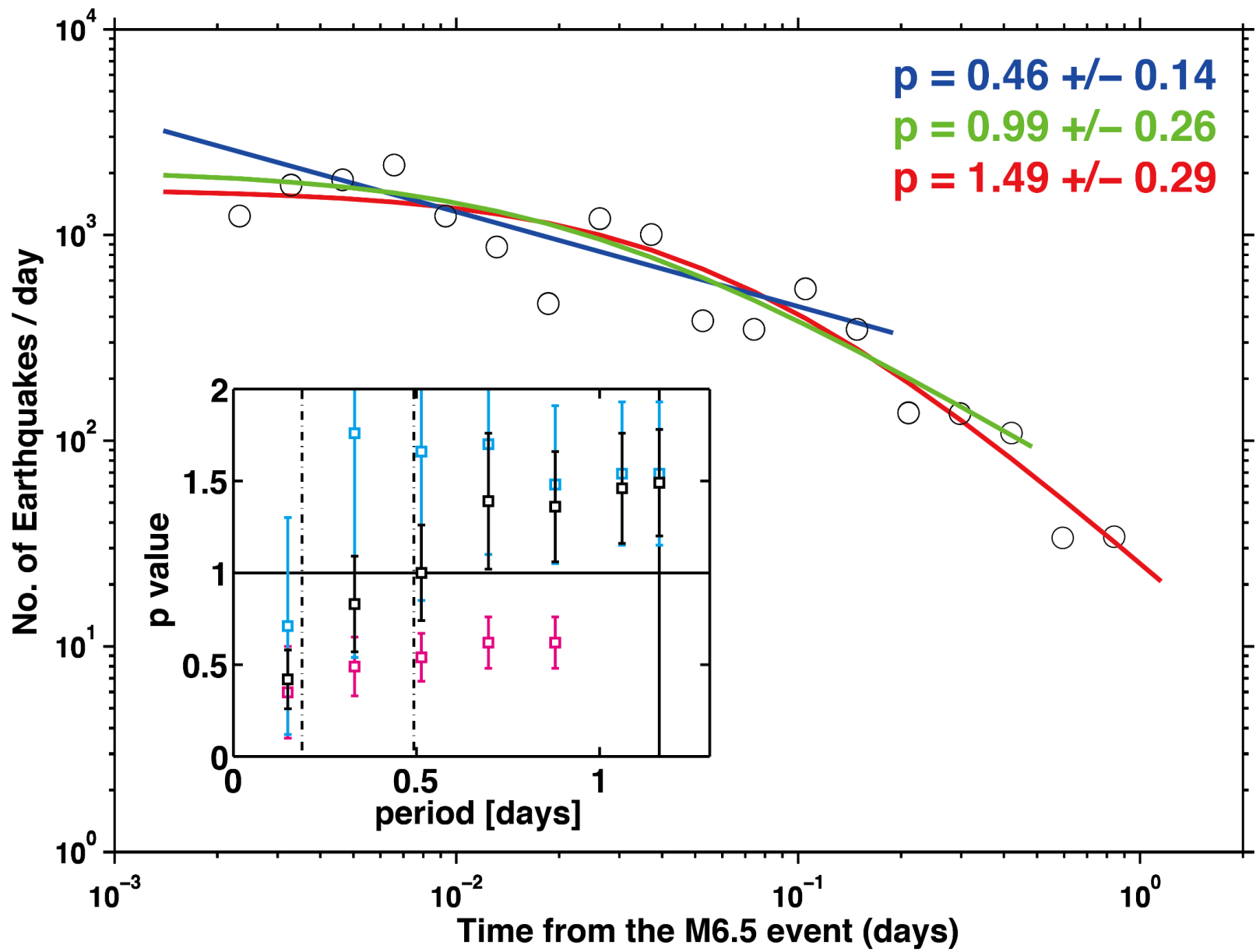

Fig. 3

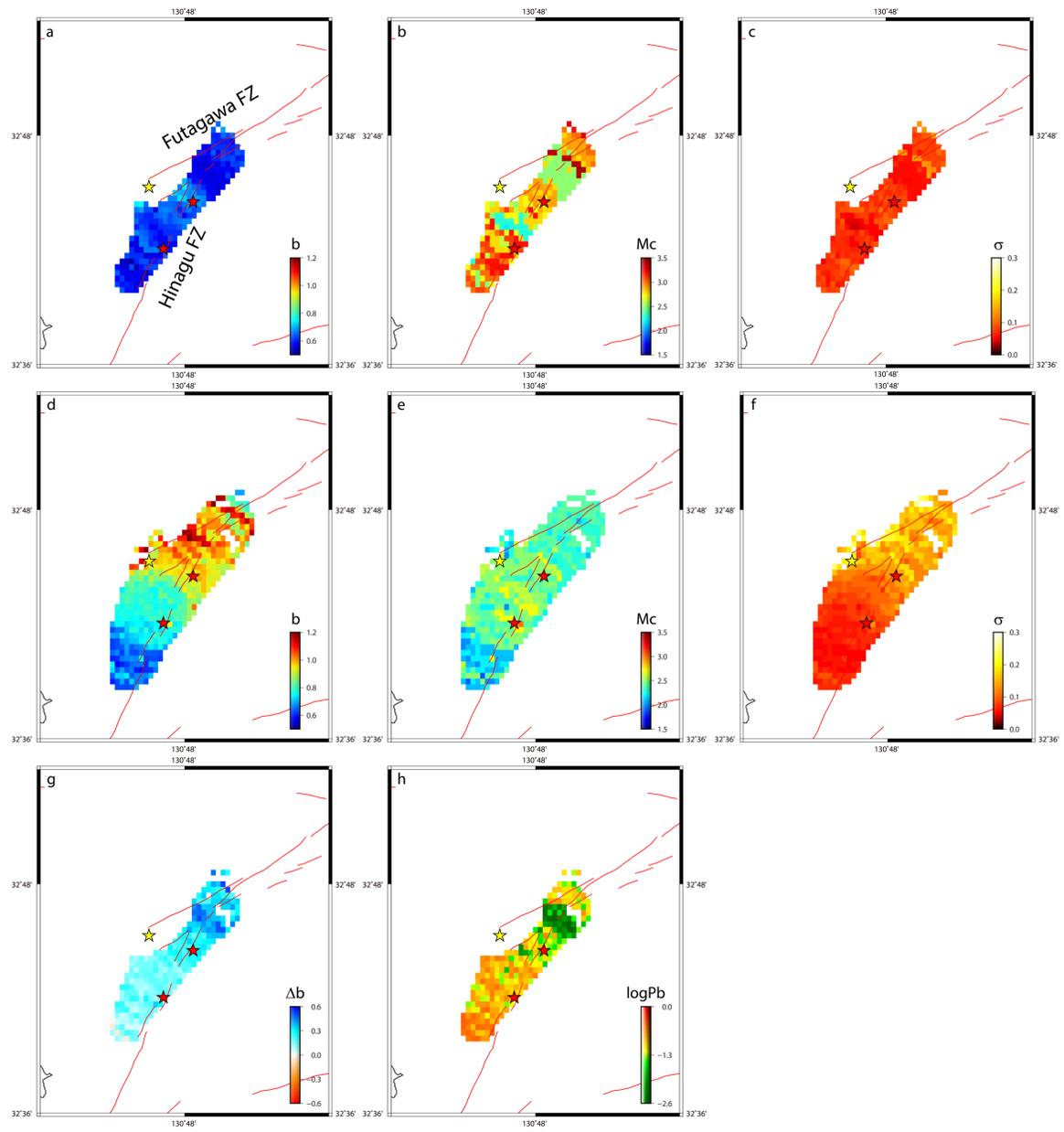

Fig. 4

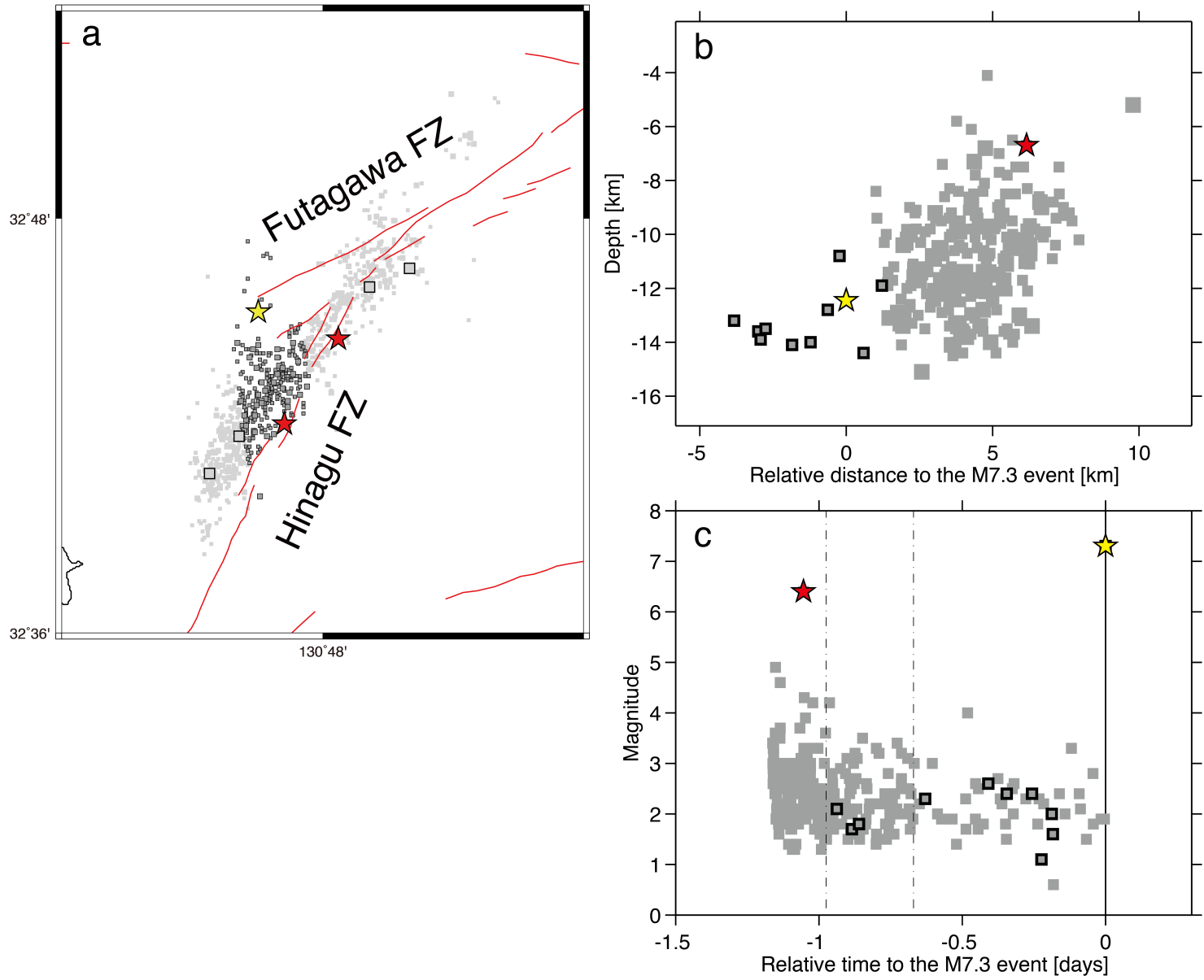

Fig. 5